\documentclass[conference]{IEEEtran}
\IEEEoverridecommandlockouts
\usepackage{amsmath,amssymb,bm}
\usepackage{graphicx}
\usepackage{xcolor}
\usepackage{booktabs}
\usepackage{microtype}
\usepackage{stfloats}
\usepackage{tikz}
\usetikzlibrary{arrows.meta,positioning,calc,fit}
\usepackage[hidelinks]{hyperref}
\usepackage{balance}
\newcommand{\ket}[1]{\lvert #1\rangle}
\newcommand{\bra}[1]{\langle #1\rvert}
\newcommand{\diag}{\operatorname{diag}}

\newcommand{\cC}{\mathcal C}
\newcommand{\Fe}{F_{\rm e}}
\newcommand{\FG}{F_{\rm GHZ}}
\definecolor{revisionred}{RGB}{195,0,0}
\ifdefined\CLEANVERSION\colorlet{revisionred}{black}\fi

\definecolor{propagationblue}{RGB}{0,0,220}
\ifdefined\CLEANVERSION\colorlet{propagationblue}{black}\fi

\begin{document}
\flushbottom

\title{Joint GKP Encoding for GHZ States Transmission over Bosonic Quantum MIMO Channels}
\IEEEoverridecommandlockouts

\author{%
\IEEEauthorblockN{Seid Koudia,  Selssabil Kadid, 
Symeon Chatzinotas, \textit{IEEE Fellow}, Mérouane Debbah, \textit{IEEE Fellow}}
\thanks{%
Seid Koudia and Symeon Chatzinotas are with the Interdisciplinary
Centre for Security, Reliability and Trust (SnT), University of
Luxembourg, Luxembourg.
% Hakim Mallil is with \'Ecole Nationale Polytechnique,
% Algiers, Algeria.
Selssabil Kadid is with \'Ecole Nationale Sup\'erieure
d'Informatique (ESI), Algiers, Algeria.
Mérouane Debbah is with the Research Institute for Digital Future,
Khalifa University of Science and Technology, Abu Dhabi, United Arab Emirates.
}}
\maketitle

\begin{abstract}
We study transmission of a Greenberger–Horne–Zeilinger (GHZ) state over nonunitary quantum multiple-input multiple-output (QuMIMO) channels using finite-energy Gottesman–Kitaev–Preskill (GKP) encoding. A coupled-mode model describes coherent mixing, mode-dependent attenuation, and Gaussian noise. The transceiver combines programmable passive meshes, quantum-limited MMSE gain, local GKP recovery, and joint classical syndrome decoding. We optimize complex-Givens lattices using instantaneous channel-state information (CSI), current singular values with channel statistics, or statistics alone. Simulations show that joint lattice shaping outperforms product GKP processing. Under coherent-basis drift, full instantaneous CSI improves GHZ-state preservation, whereas singular-value-conditioned and fully statistical designs perform similarly because neither resolves the current singular frames. These findings identify singular-frame geometry as a critical resource beyond attenuation knowledge where joint GKP processing adapts the logical lattice and receiver to anisotropic bosonic noise rather than merely inverting multimode mixing.

\end{abstract}

\section{Introduction}
Multimode optical links form quantum multiple-input multiple-output (QuMIMO) channels when spatial, temporal, frequency, or modal degrees of freedom propagate jointly. Known unitary mixing can be removed by a passive interferometer \cite{Koudia2025}, whereas coherent mixing combined with unequal attenuation is nonunitary. Therefore, inversion of the singular amplitudes requires physical gain and therefore adds quantum noise~\cite{Weedbrook2012,Caruso2008,Caves1982}. Non-Gaussian State encodings \cite{koudia2021quantum}, in particular, GKP encodings are attractive because passive multimode circuits reshape the decision lattice without increasing the number of transmitted modes, while modular syndromes support analog joint decoding~\cite{GKP2001,Conrad2022,Royer2022,Lin2023, koudia2026measurement}.

The central question is which channel information enables this lattice adaptation. Singular values characterize attenuation strengths, but do not determine the instantaneous input and output singular frames that orient the logical lattice relative to the physical channel. Our results identify this singular-frame geometry as a critical resource under coherent-basis drift. In this regime, observing current singular values supports gain retuning and conditional decoder adaptation, yet provides limited benefit when the coherent frames remain unknown. This explains why singular-value-conditioned and fully statistical designs can exhibit similar performance. Joint GKP processing therefore adapts the logical code and receiver to the anisotropic bosonic noise geometry created by nonunitary propagation.

We investigate this principle through preservation of a multipartite GHZ state over a physically parameterized multimode link. Our contributions comprise a coupled-mode Gaussian channel model with explicit propagation distance, a finite-energy GKP logical-channel description and GHZ-sector reduction, CSI-constrained stochastic compilation of optical encoders and decoders, and distance, crosstalk, and CSI benchmarks based on the complete residual logical Pauli distribution. The architecture combines joint passive optical processing with local sequential Steane recovery on each oscillator. The analog syndrome coordinates are then processed jointly by a classical coset likelihood decoder. Figure~\ref{fig:architecture} presents these stages and identifies where CSI enters the optical and classical controllers.

\section{Channel and Transceiver Model}
\subsection{Gaussian QuMIMO propagation}
Let $\hat{\bm a}(z)$ collect the three annihilation operators. A longitudinally varying passive link obeys:
\begin{align}
\frac{d\hat{\bm a}}{dz}&=K(z)\hat{\bm a}(z)+B(z)\hat{\bm e}(z),\nonumber\\
K(z)&=-\frac{\Gamma(z)}{2}-iH(z),\quad
B(z)B^\dagger(z)=\Gamma(z),
\label{eq:langevin}
\end{align}
where $H(z)=H^\dagger(z)$ is the coherent propagation matrix and $\Gamma(z)=\Gamma^\dagger(z)\succeq0$ is the attenuation/dissipative-coupling matrix. Define
\begin{equation}
T(\ell,z)=\mathcal P\exp\!\left[\int_z^\ell K(s)\,ds\right],
\qquad T(\ell)=T(\ell,0),
\label{eq:orderedpropagator}
\end{equation}
where $\mathcal P$ orders larger propagation coordinates to the left. Ordering is required when generators at different positions do not commute. The distributed-noise solution is
\begin{equation}
\hat{\bm a}(\ell)=T(\ell)\hat{\bm a}(0)+\int_0^{\ell}T(\ell,z)B(z)\hat{\bm e}(z)\,dz .
\label{eq:solution}
\end{equation}
Below we specialize to a longitudinally uniform effective link, with $H$, $\Gamma$, and $B$ constant along $z$ for each realization. Thus $K(z)=K$, $T(\ell)=e^{K\ell}$, and $T(\ell,z)=e^{K(\ell-z)}$. 

We parameterize
\begin{equation}
H_{ii}=\omega_i,\quad H_{ij}=g_{ij}e^{i\phi_{ij}},\qquad
\Gamma_{ii}=\gamma_i,\quad \Gamma_{ij}=\gamma_{ij}e^{i\psi_{ij}},
\label{eq:physicalparams}
\end{equation}
with detunings $\omega_i$, coherent coupling magnitudes/phases $(g_{ij},\phi_{ij})$, modal losses $\gamma_i$, and dissipative couplings $(\gamma_{ij},\psi_{ij})$.\footnote{These effective modal parameters can be mapped to free-space optical (FSO) link parameters through the paraxial propagation and mode-overlap construction of Peng \emph{et al.}~\cite{Peng2026FSO}. Wavelength, beam geometry, turbulence statistics and receiver apertures specify an ensemble of modal transfer matrices, from which effective coupling and loss parameters can be calibrated.} The rates $\omega_i,g_{ij},\gamma_i,\gamma_{ij}$ are specified per unit length. In the uncoupled limit, $\eta_i(\ell)=e^{-\gamma_i\ell/2}=10^{-\alpha_i\ell/20}$ for power attenuation $\alpha_i$ in dB/km. At distance $\ell$,
\begin{equation}
T(\ell)=e^{K\ell}=U_R(\ell)D_\eta(\ell)U_T^\dagger(\ell),\quad
D_\eta=\diag(\eta_1,\eta_2,\eta_3),
\label{eq:svd}
\end{equation}
where the second equality is the SVD of the same transfer matrix and $\eta_j(\ell)=\sigma_j[T(\ell)]$. Nonnormality, $[K,K^\dagger]=-i[\Gamma,H]$, generally makes both singular frames distance dependent. In quadratures, $X(\ell)=\mathcal R[T(\ell)]$ and
\begin{equation}
Y(\ell)=\tfrac12[I_6-X(\ell)X^T(\ell)]+Y_{\rm ex}(\ell),
\label{eq:Y}
\end{equation}
where $Y_{\rm ex}\succeq0$ is correlated excess noise. Equivalently, the physical channel coordinates are $x=(\bm\omega,\bm\gamma,\{g_{ij},\phi_{ij}\},\{\gamma_{ij},\psi_{ij}\},Y_{\rm ex})$. The controller descriptor is therefore the Gaussian pair $(T,Y)$ or a probability law over these coordinates.

The information sets used throughout are
\begin{align}
\mathcal I_0&=\varnothing,\qquad
\mathcal I_1=\{T(x),Y(x)\},\nonumber\\
\mathcal I_2&=\{p(T,Y),D_\eta(x)\},\qquad
\mathcal I_3=\{p(T,Y)\}.
\label{eq:infosets}
\end{align}
where statistical CSI is the complete joint law $p(T,Y)=p(U_R,D_\eta,U_T,Y)$, not only singular-value statistics \footnote{ The required CSI is assumed to be supplied by an offline
channel-estimation stage. Its accuracy depends on the type and
amount of estimation payload employed. The design of the
estimation protocol and the associated resource--accuracy
tradeoff are beyond the scope of this paper.}. 

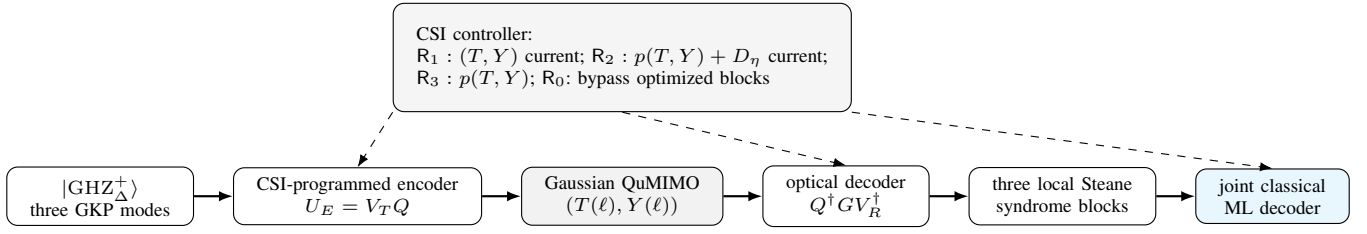
\begin{figure*}[t]
\centering
\resizebox{0.98\textwidth}{!}{%
\begin{tikzpicture}[font=\scriptsize,node distance=4mm and 5mm,
box/.style={draw,rounded corners,align=center,minimum height=7mm,inner xsep=3mm},
ctrl/.style={draw,rounded corners,align=left,fill=gray!8,inner sep=3mm},
arr/.style={-{Latex[length=1.8mm]},thick},carr/.style={-{Latex[length=1.6mm]},dashed}]
\node[box] (in) {$\ket{\mathrm{GHZ}_\Delta^+}$\\three GKP modes};
\node[box,right=of in] (enc) {CSI-programmed encoder\\$U_E=V_TQ$};
\node[box,right=of enc,fill=gray!10] (ch) {Gaussian QuMIMO\\$(T(\ell),Y(\ell))$};
\node[box,right=of ch] (dec) {optical decoder\\$Q^\dagger G V_R^\dagger$};
\node[box,right=of dec] (gkp) {three local Steane\\syndrome blocks};
\node[box,right=of gkp,fill=cyan!8] (ml) {joint classical\\ML decoder};\foreach \a/\b in {in/enc,enc/ch,ch/dec,dec/gkp,gkp/ml}{\draw[arr](\a)--(\b);}
\node[ctrl,above=7mm of ch] (csi) {CSI controller:\\$\mathsf R_1:(T,Y)$ current; $\mathsf R_2:p(T,Y)+D_\eta$ current;\\$\mathsf R_3:p(T,Y)$; $\mathsf R_0$: bypass optimized blocks};
\draw[carr] (csi.south west) -- (enc.north);
\draw[carr] (csi.south) -- (dec.north);
\draw[carr] (csi.south east) -- (ml.north);
\end{tikzpicture}}
\caption{General CSI-aware Gaussian QuMIMO-GKP architecture. CSI programs the encoder lattice/frame, receiver frame and gain, and the covariance used by the joint classical likelihood decoder. GKP syndrome extraction remains local to each oscillator; only the final logical decision is joint.}
\label{fig:architecture}
\end{figure*}

Regime $\mathsf R_0$ bypasses the optimized optical blocks and applies independent local decisions.  In $\mathsf R_1$, $V_T=U_T(x)$, $V_R=U_R(x)$, the gain is retuned from $D_\eta(x)$, and the decoder uses $C_6(x)$.  Regime $\mathsf R_2$ fixes statistically compiled frames but updates the lattice, gain, and covariance model conditionally on the measured $D_\eta(x)$.  Regime $\mathsf R_3$ uses one fixed statistical configuration.

\subsection{Finite-energy GKP states and physical equalization}
For one oscillator, $[\hat q,\hat p]=i$.  The square-lattice GKP qubit is stabilized by
$S_q=e^{i2\sqrt\pi\hat q}$ and $S_p=e^{-i2\sqrt\pi\hat p}$, with logical displacements
$\bar Z=e^{i\sqrt\pi\hat q}$ and $\bar X=e^{-i\sqrt\pi\hat p}$.  Its ideal computational states are the position combs
\begin{equation}
\langle q|b_L\rangle\propto\sum_{n\in\mathbb Z}\delta\!\left(q-(2n+b)\sqrt\pi\right),\qquad b\in\mathbb Z_2 .
\label{eq:gkpideal}
\end{equation}
Because Eq.~\eqref{eq:gkpideal} has infinite energy, we use the normalized damped state
$\ket{\widetilde b}_\Delta=\mathcal N_b e^{-\Delta^2\hat n}\ket{b_L}$, whose peaks and global envelope have finite width.  Finite-energy GKP states consist of finite-width peaks confined within a normalizing envelope. The parameter $v_{\rm env}$ characterizes the envelope's quadrature second moment. For symmetric states in the narrow-peak limit, $\langle\hat q^2\rangle\simeq\langle\hat p^2\rangle\simeq v_{\rm env}$ and the peak variance satisfies $\delta\simeq\Delta^2/2\simeq(4v_{\rm env})^{-1}$. Consequently, the mean photon number per oscillator is $\bar n=\bigl(\langle\hat q^2\rangle+\langle\hat p^2\rangle-1\bigr)/2\simeq v_{\rm env}-1/2$, while the equivalent peak squeezing relative to vacuum variance $1/2$ is $s_{\rm GKP}=-10\log_{10}(2\delta)\simeq10\log_{10}(2v_{\rm env})$ dB~\cite{GrimsmoPuri2021}. Increasing $v_{\rm env}$ therefore produces narrower GKP peaks and stronger squeezing at the cost of a broader envelope and higher oscillator energy.

Raw damped codewords need not be exactly orthogonal.  For $\bm\mu\in\mathbb Z_2^3$, set
$\ket{\widetilde{\bm\mu}}_\Delta=\bigotimes_j\ket{\widetilde{\mu_j}}_\Delta$ and
$[\mathsf G_\Delta]_{\bm\mu\bm\nu}=\langle\widetilde{\bm\mu}|\widetilde{\bm\nu}\rangle$.  Symmetric L\"owdin orthonormalization defines
\begin{equation}
\ket{\bm\mu}_{\Delta,\perp}=\sum_{\bm\nu}\ket{\widetilde{\bm\nu}}_\Delta[\mathsf G_\Delta^{-1/2}]_{\bm\nu\bm\mu},\quad
W_\Delta=\sum_{\bm\mu}\ket{\bm\mu}_{\Delta,\perp}\bra{\bm\mu}_L,
\label{eq:isometry}
\end{equation}
so $W_\Delta^\dagger W_\Delta=I_8$.  The transmitted state is
\begin{equation}
\ket{\mathrm{GHZ}_\Delta^+}=W_\Delta\frac{\ket{000}+\ket{111}}{\sqrt2},
\label{eq:ghz}
\end{equation}
an entangled state of three independently addressable logical GKP qubits, with no additional repetition code.

The transceiver is $U_E=V_TQ$ and $U_D=Q^\dagger G V_R^\dagger$, where $Q\in U(3)$ reshapes the joint GKP lattice and $G=\diag(g_1,g_2,g_3)$.  If $S_M=\mathcal R(M)$, the decoder sees
\begin{align}
A&=S_{Q^\dagger}S_GS_{V_R^\dagger}XS_{V_T}S_Q,\nonumber\\
N&=S_{Q^\dagger}\!\left[S_GS_{V_R^\dagger}YS_{V_R^\dagger}^{T}S_G^T+Y_G\right]S_{Q^\dagger}^{T},
\label{eq:AN}
\end{align}
where $Y_G=\tfrac12\diag(\bm h,\bm h)$, $h_j=|g_j^2-1|$, in the $(q_1,q_2,q_3,p_1,p_2,p_3)$ ordering.  For singular transmission $\eta_j$, the no-overshoot MMSE gain is
\begin{equation}
g_j^{\rm MMSE}=\Pi_{[1,1/\eta_j]}\!\left[\frac{\eta_jv_s}{\eta_j^2v_s+(2-\eta_j^2)/2}\right],
\label{eq:mmse}
\end{equation}
with $\Pi_{[a,b]}(u)=\min\{b,\max\{a,u\}\}$ and design signal variance $v_s$ specified for each benchmark in Sec.~IV.

\section{Logical Decoder and CSI-Constrained Optimization}
The $J_I$ and GHZ-specific CSI lattice searches use the three-complex-Givens chart \cite{Clements2016}
\begin{align}
Q(\bm\vartheta)&=G_{12}(\theta_{12},\phi_{12})G_{13}(\theta_{13},\phi_{13})
G_{23}(\theta_{23},\phi_{23}),\nonumber\\
\bm\vartheta&=(\theta_{12},\theta_{13},\theta_{23},\phi_{12},\phi_{13},\phi_{23})\in\Theta,\nonumber\\
\Theta&=[0,\pi/2]^3\times[-\pi,\pi)^3.
\label{eq:qmesh}
\end{align}
with identity action outside modes $i,j$ and active block
\begin{equation}
[G_{ij}(\theta,\phi)]_{ij}=\begin{bmatrix}
\cos\theta&-e^{-i\phi}\sin\theta\\ e^{i\phi}\sin\theta&\cos\theta
\end{bmatrix}.
\label{eq:givens}
\end{equation}
For the convention in Eq.~\eqref{eq:givens}, the coupled-mode power transmissivity and reflectivity are $\cos^2\theta$ and $\sin^2\theta$, while $\phi$ is the relative optical phase.  A balanced-MZI realization therefore implements each $G_{ij}$ with two fixed 50:50 couplers and programmable internal/external phase shifts.   The following convention 
\begin{equation*}
G_{ij}(\theta,\phi)^\dagger=G_{ij}(-\theta,\phi)
=G_{ij}(\theta,\phi+\pi),
\end{equation*}
is used, with phases reduced modulo $2\pi$. Negative compiled angles are therefore equivalent to nonnegative angles with shifted phases.

Local sequential-Steane recovery produces the ordered analog syndrome
$\bm s=(s_{q1},s_{q2},s_{q3},s_{p1},s_{p2},s_{p3})^T\in[-\sqrt\pi/2,\sqrt\pi/2)^6$.  For the two recovery orders,
\begin{align}
C_{q\to p}&=C_{\rm phys}+S_Q^T\diag(2\delta I_3,3\delta I_3)S_Q,\nonumber\\
C_{p\to q}&=C_{\rm phys}+S_Q^T\diag(3\delta I_3,2\delta I_3)S_Q,\nonumber\\
C_{\rm phys}&=(A-I_6)v_{\rm env}(A-I_6)^T+N.
\label{eq:cov}
\end{align}
A parity vector $e=(\bm x,\bm z)\in\mathbb Z_2^6$ labels the logical Pauli $P_e=X^{\bm x}Z^{\bm z}$.  Its wrapped Gaussian likelihood is
\begin{align}
\mathcal L_e(\bm s|C)&=\frac{1}{(2\pi)^3\sqrt{\det C}}\sum_{\bm k\in\mathbb Z^6}
\exp\!\left[-\tfrac12\bm y_{e,\bm k}^TC^{-1}\bm y_{e,\bm k}\right],\nonumber\\
\bm y_{e,\bm k}&=\bm s-\sqrt\pi(e+2\bm k),\qquad
\hat e(\bm s)=\arg\max_{e\in\mathbb Z_2^6}\mathcal L_e(\bm s|C).
\label{eq:likelihood}
\end{align}
The normalization is common to all cosets for fixed $C$.  Let $f_e^{(o)}(\bm s)$ be the wrapped joint density of syndrome and true parity for recovery order $o\in\{q\to p,p\to q\}$.  The probability of residual class $r$ is
\begin{align}
p_r^{(o)}&=\sum_{e\in\mathbb Z_2^6}\int_{\mathcal V} f_e^{(o)}(\bm s)\,
\mathbf 1\!\left[e\oplus\hat e_o(\bm s)=r\right]d\bm s,\nonumber\\
\mathcal V&=[-\sqrt\pi/2,\sqrt\pi/2)^6.
\label{eq:residualintegral}
\end{align}
and $p_r=\tfrac12(p_r^{q\to p}+p_r^{p\to q})$.  %Randomized quasi-Monte Carlo estimates Eq.~\eqref{eq:residualintegral}.
Thus syndrome extraction is local to each oscillator, but the classical decision is joint over all six coordinates and all  Pauli cosets. The recovered channel is $\Lambda_L(\rho)=\sum_e p_eP_e\rho P_e^\dagger$.  For Eq.~\eqref{eq:ghz}, the  residual classes collapse into
$\ket{G_{\bm b}^{\pm}}=(\ket{\bm b}\pm\ket{\bar{\bm b}})/\sqrt2$, $\bm b\in\{000,001,010,011\}$, and
\begin{equation}
\rho_{\rm GHZ}^{\rm out}=\sum_{\bm b,\sigma=\pm}P_{\bm b,\sigma}\ket{G_{\bm b}^{\sigma}}\!\bra{G_{\bm b}^{\sigma}},\quad
P_{\bm b,\sigma}=\sum_{e:\,P_e\ket{G_{000}^+}\sim\ket{G_{\bm b}^{\sigma}}}p_e.
\label{eq:ghzdiag}
\end{equation}

The encoder/decoder objectives are functionals of the complete residual distribution.  Let $\pi=I_8/8$, $\Phi_8=\ket{\Phi_8}\!\bra{\Phi_8}$ with $\ket{\Phi_8}=8^{-1/2}\sum_{j=0}^7\ket j\ket j$, and $\rho_G=\ket{G_{000}^+}\!\bra{G_{000}^+}$.  Then
\begin{align}
J_F(\cC;x)&=\Fe[\Lambda_L]=p_0,\nonumber\\
J_I(\cC;x)&=S[\Lambda_L(\pi)]-S[(I\otimes\Lambda_L)(\Phi_8)]\nonumber\\
&=3-H_2(\bm p),\nonumber\\
J_G(\cC;x)&=\mathrm{Tr}[\rho_G\Lambda_L(\rho_G)]\nonumber=\sum_{P_e\in\mathcal S_{\rm GHZ}}p_e.
\end{align}
Here $H_2$ is base-2 entropy and $\mathcal S_{\rm GHZ}=\langle XXX,ZZI,IZZ\rangle$ \cite{GuhneToth2009}. Thus $J_F$ rewards only the identity residual class, $J_I=I_c^{(\pi)}$ rewards low entropy of all  classes, and $J_G$ rewards all eight GHZ-stabilizer classes. Maximizing one objective, therefore, need not maximize the others.  We denote $J_G=\FG$ for which $\FG>1/2$ certifies genuine tripartite entanglement~\cite{GuhneToth2009}.

For a known realization $x=(T,Y)$, the channel-level lattices solve
$\bm\vartheta_J^*(x)=\arg\max_{\bm\vartheta}J[\cC(\bm\vartheta);x]$, $J\in\{J_F,J_I\}$.  These meshes are then evaluated on the GHZ state without changing their parameters, testing whether generic logical-channel optimization preserves a specific multipartite input.  For the GHZ-specific CSI comparison, with
$\cC=(V_T,Q,G,V_R,\mathcal D_C)$,
\begin{align}
\label{eq:csicontrol}
\cC_1^*(x)&=\arg\max_{\cC}J_G(\cC;x),\nonumber\\
\cC_2^*(d)&=\arg\max_{\cC}\mathbb E_X[J_G(\cC;X)\mid D_\eta=d],\nonumber\\
\cC_3^*&=\arg\max_{\cC}\mathbb E_X[J_G(\cC;X)].
\end{align}
The optical admissible sets are
\begin{align}
\mathcal A_1(x):\quad &(V_T,V_R)=(U_T(x),U_R(x)),\nonumber\\[-1mm]
&(G,\mathcal D_C)=(G[D_\eta(x)],\mathcal D_{C_6(x)}),\nonumber\\
\mathcal A_2(d):\quad &(V_T,V_R)=(V_T^{\rm stat},V_R^{\rm stat}),\nonumber\\[-1mm]
&(G,\mathcal D_C)=(G[d],\mathcal D_{\bar C_6(d)}),\nonumber\\
\mathcal A_3:\quad &(V_T,V_R)=(V_T^{\rm stat},V_R^{\rm stat}),\nonumber\\[-1mm]
&(G,\mathcal D_C)=(G^{\rm stat},\mathcal D_{\bar C_6}).
\label{eq:admissible}
\end{align}
with $Q$ optimized inside the corresponding set.  For ensemble samples $x_n\sim p(T,Y)$, the statistical compiler uses the empirical functional
\begin{align}
\widehat J_3(\cC)&=\frac1N\sum_{n=1}^N J_G(\cC;x_n),\nonumber\\
\widehat J_2(\cC;d)&=\frac1{N_d}\sum_{n\in\mathcal B(d)}J_G(\cC;x_n).
\label{eq:empirical}
\end{align}
where $\mathcal B(d)$ is the conditional training bin for the measured singular spectrum.  The corresponding statistical decoder covariances are
$\bar C_6(d)=\mathbb E[C_6(X,Q_2^*(d))\mid D_\eta=d]$ and
$\bar C_6=\mathbb E[C_6(X,Q_3^*)]$.  Hence $\mathsf R_2$ updates only components identifiable from the current singular spectrum, whereas $\mathsf R_3$ fixes the complete program from $p(T,Y)$.

\subsection{Implementation and complexity}
A generic $U(3)$ block requires at most three two-mode Givens couplers and three output phases, so optical depth is fixed with distance. The regimes change the programmed parameters and update rate, not the component count.

With the shell $\mathcal K_R=\{\bm k\in\mathbb Z^6:\|\bm k\|_\infty\le R\}$, the wrapped sum contains $N_{\rm sh}=(2R+1)^6$ representatives and direct evaluation scales as $O(64N_{\rm sh})$ quadratic forms per syndrome.  In practice, representatives are pruned by Mahalanobis radius and accumulated with log-sum-exp arithmetic.  The inverse covariances and offsets can be precomputed for $C_6(x)$, $\bar C_6(d)$, or $\bar C_6$.

\section{Numerical Results}
\begin{figure*}[t]
\centering
\includegraphics[width=0.82\textwidth]{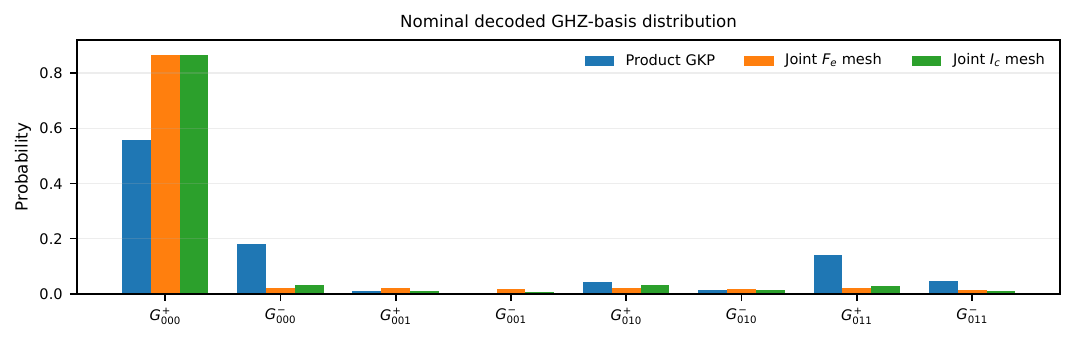}
\caption{Reference-spectrum GHZ-sector probabilities for the product baseline and $J_F$ and $J_I$ optimized channel meshes of Table~\ref{tab:nominal}}
\label{fig:nominalsectors}
\end{figure*}

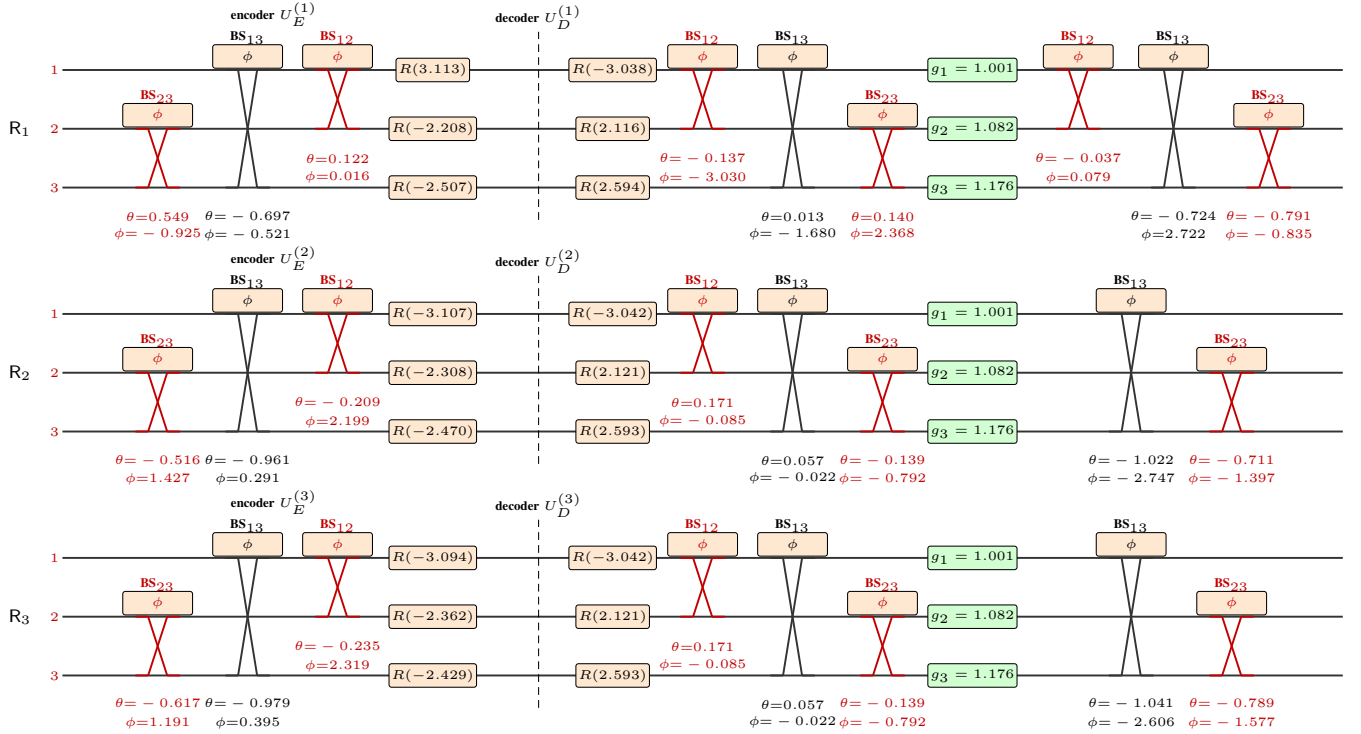
\begin{figure*}[t]
\centering
\resizebox{0.98\textwidth}{!}{%
\begin{tikzpicture}[x=0.72cm,y=0.80cm,
wire/.style={line width=0.75pt,draw=black!80},
rowlab/.style={font=\bfseries\scriptsize,align=right},
head/.style={font=\tiny\bfseries,align=center},
par/.style={font=\tiny,align=center},
phase/.style={draw,rounded corners=1pt,fill=orange!18,minimum width=0.95cm,minimum height=0.32cm,inner sep=1pt,font=\tiny},
gain/.style={draw,rounded corners=1pt,fill=green!18,minimum width=0.95cm,minimum height=0.32cm,inner sep=1pt,font=\tiny}]
\newcommand{\bsTT}[4]{\begin{scope}[wire/.style={line width=0.75pt,draw=revisionred},every node/.append style={text=revisionred}]%
\draw[wire] (#1-0.42,2)--(#1-0.18,2); \draw[wire] (#1-0.42,1)--(#1-0.18,1);
\draw[wire] (#1-0.18,2)--(#1+0.18,1); \draw[wire] (#1-0.18,1)--(#1+0.18,2);
\draw[wire] (#1+0.18,2)--(#1+0.42,2); \draw[wire] (#1+0.18,1)--(#1+0.42,1);
\node[phase] at (#1,2.23) {$\phi$}; \node[head] at (#1,2.55) {#2};
\node[par] at (#1,0.33) {\shortstack{$\theta{=}#3$\\$\phi{=}#4$}};\end{scope}}
\newcommand{\bsOT}[4]{%
\draw[wire] (#1-0.42,2)--(#1-0.18,2); \draw[wire] (#1-0.42,0)--(#1-0.18,0);
\draw[wire] (#1-0.18,2)--(#1+0.18,0); \draw[wire] (#1-0.18,0)--(#1+0.18,2);
\draw[wire] (#1+0.18,2)--(#1+0.42,2); \draw[wire] (#1+0.18,0)--(#1+0.42,0);
\node[phase] at (#1,2.23) {$\phi$}; \node[head] at (#1,2.55) {#2};
\node[par] at (#1,-0.65) {\shortstack{$\theta{=}#3$\\$\phi{=}#4$}};}
\newcommand{\bsOO}[4]{\begin{scope}[wire/.style={line width=0.75pt,draw=revisionred},every node/.append style={text=revisionred}]%
\draw[wire] (#1-0.42,1)--(#1-0.18,1); \draw[wire] (#1-0.42,0)--(#1-0.18,0);
\draw[wire] (#1-0.18,1)--(#1+0.18,0); \draw[wire] (#1-0.18,0)--(#1+0.18,1);
\draw[wire] (#1+0.18,1)--(#1+0.42,1); \draw[wire] (#1+0.18,0)--(#1+0.42,0);
\node[phase] at (#1,1.23) {$\phi$}; \node[head] at (#1,1.55) {#2};
\node[par] at (#1,-0.65) {\shortstack{$\theta{=}#3$\\$\phi{=}#4$}};\end{scope}}

% R1
\node[rowlab] at (-0.8,1) {$\mathsf R_1$};
\foreach \y in {2,1,0}{\draw[wire] (0,\y)--(24.2,\y);}
\foreach \y/\m in {2/1,1/2,0/3}{\node[font=\tiny,text=revisionred] at (-0.15,\y) {$\m$};}
\node[head] at (4.0,2.95) {encoder $U_E^{(1)}$};
\bsOO{1.8}{BS$_{23}$}{0.549}{-0.925}
\bsOT{3.5}{BS$_{13}$}{-0.697}{-0.521}
\bsTT{5.2}{BS$_{12}$}{0.122}{0.016}
\node[phase] at (7.0,2) {$R(3.113)$}; \node[phase] at (7.0,1) {$R(-2.208)$}; \node[phase] at (7.0,0) {$R(-2.507)$};
\draw[densely dashed] (9.0,-0.55)--(9.0,2.65); \node[head] at (9.0,2.90) {decoder $U_D^{(1)}$};
\node[phase] at (10.4,2) {$R(-3.038)$}; \node[phase] at (10.4,1) {$R(2.116)$}; \node[phase] at (10.4,0) {$R(2.594)$};
\bsTT{12.1}{BS$_{12}$}{-0.137}{-3.030}
\bsOT{13.8}{BS$_{13}$}{0.013}{-1.680}
\bsOO{15.5}{BS$_{23}$}{0.140}{2.368}
\node[gain] at (17.2,2) {$g_1=1.001$}; \node[gain] at (17.2,1) {$g_2=1.082$}; \node[gain] at (17.2,0) {$g_3=1.176$};
\bsTT{19.2}{BS$_{12}$}{-0.037}{0.079}
\bsOT{21.0}{BS$_{13}$}{-0.724}{2.722}
\bsOO{22.8}{BS$_{23}$}{-0.791}{-0.835}

% R2
\begin{scope}[shift={(0,-4.15)}]
\node[rowlab] at (-0.8,1) {$\mathsf R_2$};
\foreach \y in {2,1,0}{\draw[wire] (0,\y)--(24.2,\y);}
\foreach \y/\m in {2/1,1/2,0/3}{\node[font=\tiny,text=revisionred] at (-0.15,\y) {$\m$};}
\node[head] at (4.0,2.95) {encoder $U_E^{(2)}$};
\bsOO{1.8}{BS$_{23}$}{-0.516}{1.427}
\bsOT{3.5}{BS$_{13}$}{-0.961}{0.291}
\bsTT{5.2}{BS$_{12}$}{-0.209}{2.199}
\node[phase] at (7.0,2) {$R(-3.107)$}; \node[phase] at (7.0,1) {$R(-2.308)$}; \node[phase] at (7.0,0) {$R(-2.470)$};
\draw[densely dashed] (9.0,-0.55)--(9.0,2.65); \node[head] at (9.0,2.90) {decoder $U_D^{(2)}$};
\node[phase] at (10.4,2) {$R(-3.042)$}; \node[phase] at (10.4,1) {$R(2.121)$}; \node[phase] at (10.4,0) {$R(2.593)$};
\bsTT{12.1}{BS$_{12}$}{0.171}{-0.085}
\bsOT{13.8}{BS$_{13}$}{0.057}{-0.022}
\bsOO{15.5}{BS$_{23}$}{-0.139}{-0.792}
\node[gain] at (17.2,2) {$g_1=1.001$}; \node[gain] at (17.2,1) {$g_2=1.082$}; \node[gain] at (17.2,0) {$g_3=1.176$};
\bsOT{20.2}{BS$_{13}$}{-1.022}{-2.747}
\bsOO{22.1}{BS$_{23}$}{-0.711}{-1.397}
\end{scope}

% R3
\begin{scope}[shift={(0,-8.30)}]
\node[rowlab] at (-0.8,1) {$\mathsf R_3$};
\foreach \y in {2,1,0}{\draw[wire] (0,\y)--(24.2,\y);}
\foreach \y/\m in {2/1,1/2,0/3}{\node[font=\tiny,text=revisionred] at (-0.15,\y) {$\m$};}
\node[head] at (4.0,2.95) {encoder $U_E^{(3)}$};
\bsOO{1.8}{BS$_{23}$}{-0.617}{1.191}
\bsOT{3.5}{BS$_{13}$}{-0.979}{0.395}
\bsTT{5.2}{BS$_{12}$}{-0.235}{2.319}
\node[phase] at (7.0,2) {$R(-3.094)$}; \node[phase] at (7.0,1) {$R(-2.362)$}; \node[phase] at (7.0,0) {$R(-2.429)$};
\draw[densely dashed] (9.0,-0.55)--(9.0,2.65); \node[head] at (9.0,2.90) {decoder $U_D^{(3)}$};
\node[phase] at (10.4,2) {$R(-3.042)$}; \node[phase] at (10.4,1) {$R(2.121)$}; \node[phase] at (10.4,0) {$R(2.593)$};
\bsTT{12.1}{BS$_{12}$}{0.171}{-0.085}
\bsOT{13.8}{BS$_{13}$}{0.057}{-0.022}
\bsOO{15.5}{BS$_{23}$}{-0.139}{-0.792}
\node[gain] at (17.2,2) {$g_1=1.001$}; \node[gain] at (17.2,1) {$g_2=1.082$}; \node[gain] at (17.2,0) {$g_3=1.176$};
\bsOT{20.2}{BS$_{13}$}{-1.041}{-2.606}
\bsOO{22.1}{BS$_{23}$}{-0.789}{-1.577}
\end{scope}
\end{tikzpicture}}
\caption{$J_G$ optimized CSI encoder and decoder circuits at the common comparison channel. Modes 1--3 run from top to bottom and gates act left-to-right. Signed $(\theta,\phi)$ values in radians describe the compiled full meshes and inverse gates, not the search coordinates of Eq.~\eqref{eq:qmesh}. Orange boxes are phases and green boxes are gains.}
\label{fig:circuits}
\end{figure*}

\begin{figure*}[t]
\centering
\includegraphics[width=\textwidth]{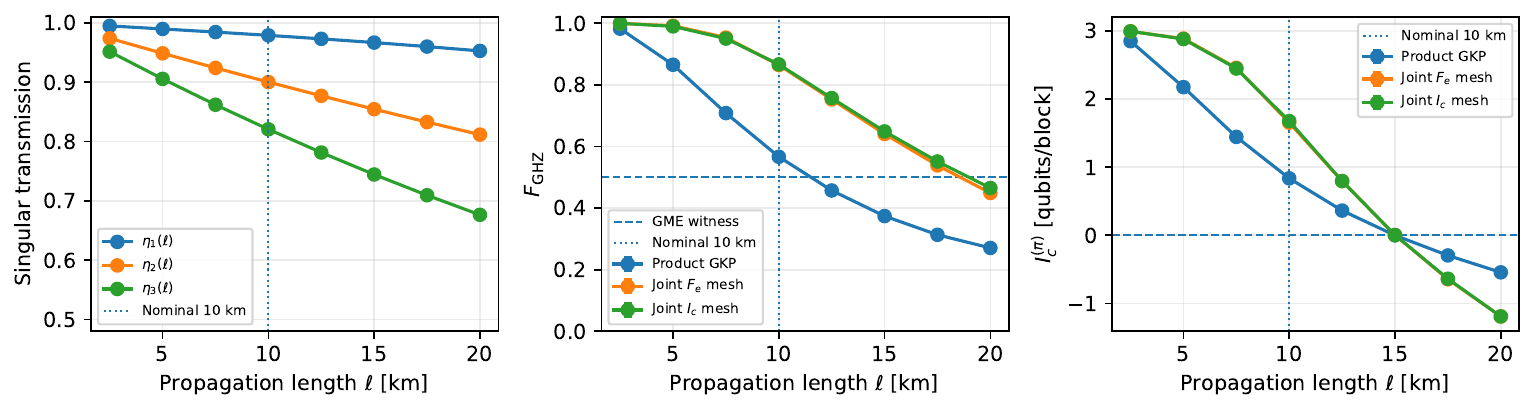}
\caption{Distance sweep with fixed reference $J_F$ and $J_I$ meshes and distance-retuned singular frames and MMSE gains. The vertical line marks the nominal 10-km channel; the horizontal lines mark the GHZ witness threshold and zero coherent information.}
\label{fig:length}
\end{figure*}

\textit{Optimization objectives and reported metrics.} Table~\ref{tab:nominal} and Fig.~\ref{fig:nominalsectors} evaluate the reference $J_F$ and $J_I$ optimized channel meshes on a GHZ input. Figure~\ref{fig:length} keeps these same $Q$ matrices fixed while updating the channel frames and gains with distance. Figure~\ref{fig:circuits} and Table~\ref{tab:csi} instead use separately $J_G$ optimized CSI designs from Eq.~\eqref{eq:csicontrol}. 

The simulations use $\bm\omega=(.010,-.005,.002)$ rad/km, $\bm g=(.035,.010,.028)$ rad/km, $\bm\phi=(.55,.80,-.35)$ rad, $\bm\gamma=(.00404,.02107,.03969)$ km$^{-1}$, $\gamma_{ij}=0$, and $Y_{\rm ex}=0$. These propagation coefficients are held constant along the link, so the distance sweep uses $T(\ell)=e^{K\ell}$. At 10 km the uncoupled reference amplitudes are $(0.98,0.90,0.82)$, while the coupled channel has singular amplitudes $(0.97901,0.90023,0.82062)$. The results in Table~\ref{tab:nominal} and Fig.~\ref{fig:nominalsectors} uses the former spectrum and fixed gain $(1,1.07175,1.16215)$. The distance sweep uses the actual coupled spectrum and Eq.~\eqref{eq:mmse} with $v_s=v_{\rm env}$.

All designs use $v_{\rm env}\simeq29.96$, giving $\delta=8.346\times10^{-3}$, $s_{\rm GKP}=17.78$ dB and $\bar n\simeq29.46$ photons per oscillator. This common preparation point minimizes the product of the two Steane-order decision-covariance determinants in the reference fixed-gain calibration. It balances decreasing peak noise, proportional to $1/v_{\rm env}$, against increasing residual-mismatch noise, proportional to $v_{\rm env}$. It is held fixed for fair comparisons, not reoptimized for every channel or CSI regime.

Table~\ref{tab:nominal} compares product processing with channel-level joint meshes at the reference spectrum and fixed gain. Joint lattice shaping concentrates the logical distribution more strongly and preserves the GHZ state more effectively, while the two channel-level objectives produce similar state fidelity through different residual-error spectra.
\begin{table}[t]
\centering
\caption{Reference-spectrum GHZ benchmark: fixed gain, $J_F$ and $J_I$ optimized channel meshes.}
\label{tab:nominal}
\setlength{\tabcolsep}{4pt}
\begin{tabular}{lccc}
\toprule
Design & $\FG$ & $\Fe$ & $I_c^{(\pi)}$\\
\midrule
Product GKP & 0.5588 & 0.5480 & 0.8005\\
Joint $F_e$ mesh & 0.8659 & 0.8494 & 1.6466\\
Joint $I_c$ mesh & \textbf{0.8666} & 0.8455 & \textbf{1.6602}\\
\bottomrule
\end{tabular}
\end{table}

Figure~\ref{fig:nominalsectors} resolves the nominal scalar metrics into GHZ error sectors. Product processing leaves substantial phase-flipped and complementary-bit weight, whereas both joint meshes concentrate probability near the target sector while differing in the distribution of the remaining logical errors.

Figure~\ref{fig:circuits} gives only the optimized encoder and optical decoder circuits. The common Gaussian channel and GKP recovery blocks are shown once in Fig.~\ref{fig:architecture}, so the circuit comparison focuses on the CSI-dependent optical hardware and parameters.
 The compiled circuits expose the CSI constraint: $\mathsf R_1$ tracks the current singular frames, while $\mathsf R_2$ and $\mathsf R_3$ share fixed statistical frames and use different lattice programs. 

Figure~\ref{fig:length} shows the coupled evolution of the singular spectrum and logical metrics. The joint meshes retain the GHZ witness over a substantially longer direct-link interval than product processing. Coherent information is more stringent: it vanishes when the residual Pauli distribution becomes too entropic even while a particular GHZ state may still retain nonzero fidelity.

\textit{Instantaneous frames, beyond attenuation knowledge.}
The CSI benchmark separates frame and loss uncertainty as
$T(\xi,\lambda)=U_R(\xi)D_\eta(\lambda)U_T^\dagger(\xi)$ with
$p(\xi,\lambda)=p(\xi)p(\lambda)$. Consequently,
$p(U_T,U_R\mid D_\eta)=p(U_T,U_R)$: observing current attenuation leaves the hidden-frame uncertainty unchanged. At the common realization $(\xi,\lambda)=(0.3,0)$, Table~\ref{tab:csi} gives $\FG=0.8701$ for $\mathsf R_1$, versus 0.4516 for $\mathsf R_2$ and 0.4573 for $\mathsf R_3$. Only $\mathsf R_1$ crosses the GHZ witness threshold there.

The near equality of $\mathsf R_2$ and $\mathsf R_3$ identifies a regime where frame mismatch dominates the benefit of gain and likelihood retuning. The corresponding mean fidelities over the declared $5\times3$ grid are approximately 0.8719, 0.4163, and 0.4166, consistent with the same interpretation.

\begin{table}[t]
\centering
\caption{Common-channel comparison of $J_G$ optimized CSI designs, $(\xi,\lambda)=(0.3,0)$; $\mathsf R_0$ is the unoptimized baseline.}
\label{tab:csi}
\setlength{\tabcolsep}{3.6pt}
\begin{tabular}{lccc}
\toprule
Regime & $\FG$ & $\Fe$ & $I_c^{(\pi)}$\\
\midrule
$\mathsf R_0$: no CSI & 0.1308 & 0.0167 & -2.9969\\
$\mathsf R_1$: full instantaneous & 0.8701 & 0.8378 & 1.6399\\
$\mathsf R_2$: $D_\eta+$ statistics & 0.4516 & 0.3565 & -0.9915\\
$\mathsf R_3$: full statistical & 0.4573 & 0.3605 & -1.0304\\
\bottomrule
\end{tabular}
\end{table}

\section{Discussion}
The central result is that instantaneous singular-frame geometry is an operational resource for GKP transmission over bosonic QuMIMO channels with coherent crosstalk. For the longitudinally uniform benchmark, the transfer $T(\ell)=e^{K\ell}$ combines coherent mode conversion with mode-dependent attenuation; when $[H,\Gamma]\neq0$, it is generally nonnormal, and its input and output singular frames can evolve differently with distance. Knowledge of the singular values specifies attenuation strengths, but does not identify the frame alignment needed to orient the logical lattice relative to the physical channel.

The physical mechanism is visible in the residual covariance,
\begin{equation*}
C_{\rm phys}
=
v_{\rm env}(A-I_6)(A-I_6)^T+N.
\end{equation*}
With current $U_T$ and $U_R$, matched encoding and receiver processing remove deterministic coherent mixing, while unequal attenuation and physical gain generally leave anisotropic residual noise. The lattice mesh changes how this noise intersects the logical decision boundaries. With mismatched frames, residual coherent mixing also contributes to $A-I_6$, and the broad finite-energy GKP envelope amplifies the resulting displacement uncertainty. Joint GKP processing is therefore not simply quantum MIMO inversion, it is rather an adaptation of the logical code and receiver to the anisotropic bosonic noise geometry created by nonunitary multimode propagation.

Under the benchmark's independent frame-and-loss prior, observing $D_\eta$ supports MMSE retuning and conditional lattice and likelihood adaptation but does not reveal $U_T$ or $U_R$. When frame mismatch dominates, this explains the similar performance of $\mathsf R_2$ and $\mathsf R_3$. Their similarity is not universal. Singular values can become more informative when frame and loss fluctuations are correlated or attenuation uncertainty dominates. The GHZ and channel metrics describe different consequences of the same logical noise. Entanglement fidelity $F_e$ rewards the identity residual class, whereas $F_{\rm GHZ}$ also benefits from nonidentity residual Paulis in $\mathcal S_{\rm GHZ}$ that preserve the target state. In contrast, $I_c^{(\pi)}$ depends on the entropy of the complete logical Pauli distribution. Genuine multipartite entanglement can therefore survive even when this single-letter coherent-information benchmark becomes nonpositive. Reporting these metrics together distinguishes preservation of a specific entangled state from performance on generic logical inputs. This adaptation retains a local quantum recovery structure. Following the passive optical decoder, each oscillator undergoes its own sequential Steane recovery. The analog syndrome outcomes are then combined by the joint classical coset decoder. Exploiting multimode correlations therefore does not require an additional joint non-Gaussian recovery operation across the data oscillators.

% \paragraph{Scope and reproducibility.}
% The benchmark sets $Y_{\rm ex}=0$ to isolate the effects of loss and coherent mixing. General Gaussian excess noise enters through Eq.~\eqref{eq:Y}, propagates through Eq.~\eqref{eq:AN}, and modifies the decision covariance and wrapped likelihood. The distance study recomputes $T(\ell)$, its singular frames, physical noise, and MMSE gains while keeping the nominal channel-level lattices and preparation energy fixed. It therefore assesses propagation robustness over an elementary direct link. All designs use the same finite-energy second-moment model, averaging over the sequential Steane recovery orders, truncated wrapped sums, and randomized Monte Carlo estimation.

\paragraph{Scaling and extensions.}
For $M$ modes, the passive optical layer requires $O(M^2)$ interferometer elements, while modewise gain and local recovery scale linearly with $M$. Exact classical Pauli-coset decoding involves $4^M$ classes, motivating structured lattices and approximate decoding for larger systems. Geometry-aware circuit optimization and repeater-assisted operation remain extensions whose performance requires separate investigation.

\section{Conclusion}
We developed a finite-energy GKP transceiver for GHZ-state transmission over bosonic QuMIMO channels, combining a physical propagation model, CSI-dependent passive meshes, quantum-limited MMSE gain, local Steane recovery, and joint classical coset decoding. The model accounts for coherent coupling, mode-dependent attenuation, and Gaussian noise. The resulting logical Pauli distribution determines GHZ-state fidelity, entanglement fidelity, and coherent information, distinguishing state preservation from generic logical-channel performance. Joint lattice shaping outperforms product-GKP processing over the investigated direct-link range. Under the independent frame-and-loss prior, singular values support gain retuning and conditional lattice and decoder adaptation but leave the instantaneous coherent basis unresolved. This explains the similar performance of singular-value-conditioned and fully statistical designs when frame mismatch dominates, whereas full instantaneous CSI enables better GHZ-state preservation. Instantaneous singular-frame geometry is therefore a critical resource. Joint GKP processing adapts the logical code and receiver to the anisotropic bosonic noise geometry created by nonunitary multimode propagation.

\bibliographystyle{IEEEtran}
\bibliography{references.bib}
\end{document}